\documentclass{article}
\usepackage{spconf,amsmath,graphicx,hyperref}
\usepackage{booktabs}

\title{GAP-URGENet: A Generative-Predictive Fusion Framework for Universal Speech Enhancement}
\name{Xiaobin Rong$^{1,2,\dagger}$, Yushi Wang$^{1,2,\dagger}$, Zheng Wang$^{1,2}$, Jing Lu$^{1,2}$ \thanks{$^\dagger$Equal contribution}}
\address{$^1$Key Laboratory of Modern Acoustics, Nanjing University \\
        $^2$NJU-Horizon Intelligent Audio Lab, Horizon Robotics \\
         \small\texttt{\{xiaobin.rong, yushi.wang, zheng.wang\}@smail.nju.edu.cn, lujing@nju.edu.cn}}
        
\begin{document}
%\ninept
%
\maketitle
\begin{abstract}
We introduce GAP-URGENet, a generative-predictive fusion framework developed for Track 1 of the ICASSP 2026 URGENT Challenge. The system integrates a generative branch, which performs full-stack speech restoration in a self-supervised representation domain and reconstructs the waveform via a neural vocoder, along with a predictive branch that performs spectrogram-domain enhancement, providing complementary cues. Outputs from both branches are fused by a post-processing module, which also performs bandwidth extension to generate the enhanced waveform at 48 kHz, later downsampled to the original sampling rate. This generative-predictive fusion improves robustness and perceptual quality, achieving top performance in the blind-test phase and ranking 1st in the objective evaluation. 
Audio examples are available at \urlstyle{same}\url{https://xiaobin-rong.github.io/gap-urgenet_demo}.
\end{abstract}
\begin{keywords}
speech enhancement, URGENT challenge, generative model, predictive model, fusion
\end{keywords}
\vspace{-5pt}
\section{Introduction}
\vspace{-10pt}
\label{sec:intro}
The URGENT Challenge \cite{URGENT2024, URGENT2025} targets universal speech enhancement (USE) under diverse distortions and sampling rates. The ICASSP 2026 edition \cite{URGENT2026} further extends this setting in Track 1 by emphasizing robustness to diverse speech sources and multiple language families.

Previous challenge results highlight the potential of combining generative and predictive methods \cite{URGENT2025}. Predictive approaches \cite{URGENT2025_rank1_tencent} achieve strong objective scores but often yield suboptimal perceptual quality, whereas purely generative models deliver better subjective performance at the risk of hallucinations. Generative-predictive fusion systems \cite{URGENT2025_rank2_nju, URGENT2025_rank3_ut} provide a better balance between objective and subjective metrics, though substantial room for improvement remains.

To address this challenge, we propose GAP-URGENet, a Generative-Allied-Predictive Universal, Robust, and Generalizable speech Enhancement Network. The system has two parallel 16-kHz branches: a generative branch that extends our earlier low-hallucination paradigm PASE \cite{PASE} to full-stack restoration, and a predictive branch that provides complementary cues to preserve faithful signal details. Outputs from both branches are fused by a post-processing module, which also performs bandwidth extension (BWE) to generate the enhanced waveform at 48 kHz before downsampling to the original rate. Results from the ICASSP 2026 URGENT Challenge demonstrate the superiority of GAP-URGENet, achieving 1st place in the objective evaluation.

\vspace{-5pt}
\section{Method}
\vspace{-10pt}
\begin{figure*}[t]
    \centering
    \includegraphics[width=0.9\linewidth]{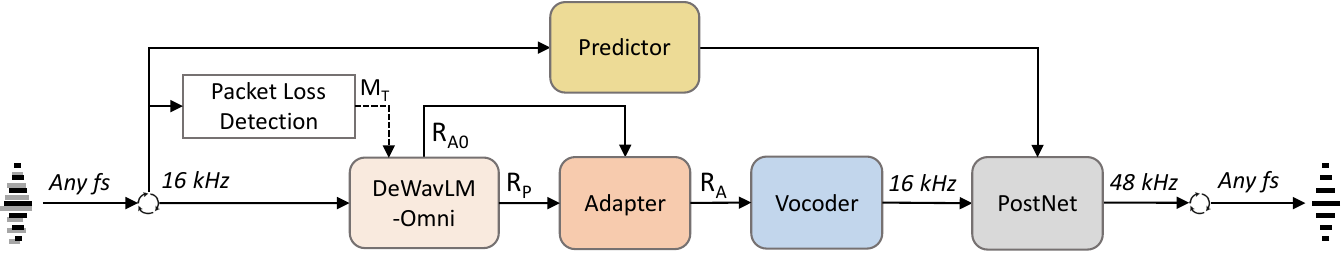}
    \vspace{-10pt}
    \caption{Overview of the GAP-URGENet framework.}
    \label{fig:gap-urgent}
\end{figure*}

As illustrated in Fig.~\ref{fig:gap-urgent}, the generative branch comprises DeWavLM-Omni, an Adapter, and a Vocoder, while the predictive branch is a Predictor. The post-processing module is the PostNet. Further details are provided below.

\begin{table*}[t]
\vspace{-15pt}
\centering
\caption{Experimental results on the official validation set.}
\label{tab:exp}
\small
\begin{tabular}{@{}cccccccccc@{}}
\toprule
Model & DNSMOS $\uparrow$ & NISQA $\uparrow$ & UTMOS $\uparrow$ & SCOREQ $\uparrow$ & PESQ $\uparrow$ & ESTOI $\uparrow$ & SBS $\uparrow$ & LPS $\uparrow$ & SpkSim $\uparrow$ \\ \midrule
Noisy & 1.80 & 1.56 & 1.58 & 1.72 & 1.26 & 0.61 & 0.70 & 0.58 & 0.58 \\
BSRNN & 3.13 & 3.48 & 2.70 & 3.32 & 2.57 & 0.85 & 0.88 & 0.82 & 0.78 \\
BSRNN-Flow & 3.19 & 3.83 & 2.65 & 3.35 & 2.15 & 0.81 & 0.83 & 0.75 & 0.74 \\ \midrule
Predictive branch & 3.20 & 3.66 & 3.13 & 3.51 & 2.64 & 0.86 & 0.86 & 0.81 & 0.70 \\
Generative branch & 3.23 & \textbf{4.01} & 3.01 & 3.97 & 2.27 & 0.83 & 0.90 & 0.87 & 0.76 \\
GAP-URGENet & \textbf{3.31} & 3.96 & \textbf{3.22} & \textbf{4.03} & \textbf{2.89} & \textbf{0.90} & \textbf{0.91} & \textbf{0.89} & \textbf{0.82} \\ \bottomrule
\end{tabular}
\vspace{-15pt}
\end{table*}

\vspace{-10pt}
\subsection{Generative Branch}
\vspace{-5pt}
We extend the core denoising module DeWavLM from PASE \cite{PASE} into a full-stack speech restoration module, forming DeWavLM-Omni. The representation-distillation paradigm from \cite{PASE} is retained, but the noisy speech is now further augmented with diverse distortions. To better exploit the masked-prediction capability of WavLM for packet loss concealment (PLC), we apply a simple packet-loss detection algorithm to identify missing frames and replace the corresponding CNN outputs in WavLM with learnable masked embeddings. 
DeWavLM-Omni outputs dual-stream representations: (1) acoustic representations from the first Transformer layer, which preserve fine-grained details while retaining residual noise and distortions, and (2) phonetic representations from the final Transformer layer, which contain rich and purified phonetic information.

We employ an Adapter to map the enhanced phonetic representations ($\textrm{R}_\textrm{P}$) to enhanced acoustic representations ($\textrm{R}_\textrm{A}$), conditioned on the noisy acoustic representations ($\textrm{R}_\textrm{A0}$), with the conditioning implemented via element-wise addition. This design provides the Vocoder with purified acoustic features, enabling high-fidelity waveform reconstruction. The Adapter follows the improved Vocos architecture proposed in WavTokenizer \cite{WavTokenizer}, excluding the iSTFT head, and is trained with an MSE loss combined with representation-domain adversarial and feature-matching losses.

The Vocoder reconstructs the enhanced 16-kHz waveform from the enhanced acoustic representations ($\textrm{R}_\textrm{A}$). It is trained independently on clean speech and integrated into the system without joint fine-tuning. The Vocoder also adopts the improved Vocos architecture, and is optimized using a multi-scale Mel-spectrogram loss combined with adversarial and feature-matching losses from a multi-period discriminator \cite{HiFiGAN} and a multi-band multi-scale STFT discriminator \cite{DAC}.

\vspace{-10pt}
\subsection{Predictive Branch}
\vspace{-5pt}
The Predictor performs predictive enhancement, offering complementary cues that preserve faithful signal details for the generative pathway. Owing to its discriminative nature, this branch specifically targets noise, reverberation, and clipping. The Predictor adopts the TF-GridNet \cite{TF-GridNet} architecture and is trained using an STFT-domain loss.

\vspace{-10pt}
\subsection{PostNet}
\vspace{-5pt}
The PostNet receives the concatenated outputs from both branches as input and internally performs feature fusion and BWE to generate a 48-kHz waveform, which is then downsampled to the original rate. It follows the CWS-TF-GridNet architecture from TS-URGENet \cite{URGENT2025_rank2_nju} and is optimized with the same loss function as the Vocoder, further incorporating PESQ- and UTMOS-aware terms as described in \cite{URGENT2025_rank2_nju}.

\vspace{-5pt}
\section{Experiments}
\vspace{-5pt}
\subsection{Datasets and Experimental Setup}
\vspace{-5pt}
We use all the corpora provided by the URGENT 2026 Challenge, adopting the original Track 1 versions for datasets also included in URGENT 2025, as they contain more data than the 2026 versions. For LibriVox, LibriTTS, VCTK, MLS, and CommonVoice 19.0, we filter samples using DNSMOS P.835 (OVRL, SIG, BAK) \cite{DNSMOS-P835}, and DNSMOS P.808 \cite{DNSMOS} scores, with a common threshold of 3. No filtering is applied to the EARS corpus. For the NNCES and SeniorTalk corpora, a DPCRN \cite{DPCRN} model pretrained on the URGENT 2025 dataset is employed to clean the data. The curated speech has a total duration of around 2,473 hours. For the noise corpora, only the Free Music Archive is pre-processed using the open-source, pre-trained SC-Net\footnote{\urlstyle{same}\url{https://github.com/starrytong/SCNet}} to remove vocal components, preventing contamination of the training data. For the RIR data, we simulate an additional 10,000 samples with high RT60 values uniformly distributed between 0.6 and 1.6~s.

The DeWavLM-Omni adopts the WavLM-Large \cite{WavLM} configuration. Both the Adapter and Vocoder share the same setup: a hidden dimension of 1024 and 12 ConvNeXt blocks with a shared intermediate dimension of 3072. The iSTFT uses an FFT size of 1280 with a hop size of 320. The Predictor and PostNet follow the configurations of TF-GridNet-L and CWS-TF-GridNet, respectively, as described in \cite{URGENT2025_rank2_nju}. The overall framework contains 567.76M parameters and requires 472.84 GMACs per second.

\vspace{-10pt}
\subsection{Results and Conclusion}
\vspace{-5pt}
Experimental results on the official validation set are summarized in Table~\ref{tab:exp}. Both our predictive and generative branches achieve substantial improvements over their respective official baselines \cite{URGENT2026_baseline}, BSRNN and BSRNN-Flow. Notably, the generative branch outperforms the predictive baseline in SpkSim and LPS, demonstrating superior speaker and linguistic fidelity and highlighting its low-hallucination characteristics. The best overall performance is achieved by GAP-URGENet, which effectively integrates the strengths of both branches.

Results on the blind-test set further confirm the superiority of GAP-URGENet, achieving 1st place in the objective evaluation\footnote{\urlstyle{same}\url{https://urgent-challenge.com/competitions/15\#final_results}}. These findings validate our predictive-generative fusion strategy and establish GAP-URGENet as a state-of-the-art solution for universal speech enhancement.

\section{Acknowledgments}
This work was supported by the National Natural Science Foundation of China (Grant No. 12274221) and the Yangtze River Delta Science and Technology Innovation Community Joint Research Project (Grant No. 2024CSJGG1100).

%\begin{small}
\bibliographystyle{IEEEbib}
\bibliography{ref}

@inproceedings{DPCRN,
 author = {Xiaohuai Le and Hongsheng Chen and Kai Chen and Jing Lu},
 bibsource = {dblp computer science bibliography, https://dblp.org},
 booktitle = {Interspeech 2021},
 pages = {2811--2815},
 title = {{DPCRN: Dual-path convolution recurrent network for single channel speech enhancement}},
 year = {2021}
}

@article{TF-GridNet,
  title={{TF-GridNet}: Integrating full-and sub-band modeling for speech separation},
  author={Wang, Zhong-Qiu and Cornell, Samuele and Choi, Shukjae and Lee, Younglo and Kim, Byeong-Yeol and Watanabe, Shinji},
  journal={IEEE/ACM Transactions on Audio, Speech, and Language Processing},
  volume={31},
  pages={3221--3236},
  year={2023},
  publisher={IEEE}
}

@article{WavLM,
  title={{WavLM}: Large-scale self-supervised pre-training for full stack speech processing},
  author={Chen, Sanyuan and Wang, Chengyi and Chen, Zhengyang and Wu, Yu and Liu, Shujie and Chen, Zhuo and Li, Jinyu and Kanda, Naoyuki and Yoshioka, Takuya and Xiao, Xiong and others},
  journal={IEEE Journal of Selected Topics in Signal Processing},
  volume={16},
  number={6},
  pages={1505--1518},
  year={2022},
  publisher={IEEE}
}

@article{DAC,
  title={High-fidelity audio compression with improved {RVQGAN}},
  author={Kumar, Rithesh and Seetharaman, Prem and Luebs, Alejandro and Kumar, Ishaan and Kumar, Kundan},
  journal={Advances in Neural Information Processing Systems},
  volume={36},
  pages={27980--27993},
  year={2023}
}

@inproceedings{DNSMOS,
 author = {Chandan K. A. Reddy and
Vishak Gopal and
Ross Cutler},
 bibsource = {dblp computer science bibliography, https://dblp.org},
 booktitle = {{IEEE} International Conference on Acoustics, Speech and Signal Processing,
{ICASSP} 2021, Toronto, ON, Canada, June 6-11, 2021},
 doi = {10.1109/ICASSP39728.2021.9414878},
 pages = {6493--6497},
 publisher = {{IEEE}},
 title = {{DNSMOS}: {A} Non-Intrusive Perceptual Objective Speech Quality Metric
to Evaluate Noise Suppressors},
 year = {2021}
}

@inproceedings{DNSMOS-P835,
 author = {Chandan K. A. Reddy and
Vishak Gopal and
Ross Cutler},
 bibsource = {dblp computer science bibliography, https://dblp.org},
 booktitle = {{IEEE} International Conference on Acoustics, Speech and Signal Processing,
{ICASSP} 2022, Virtual and Singapore, 23-27 May 2022},
 doi = {10.1109/ICASSP43922.2022.9746108},
 pages = {886--890},
 publisher = {{IEEE}},
 title = {{DNSMOS} {P.835:} {A} Non-Intrusive Perceptual Objective Speech Quality
Metric to Evaluate Noise Suppressors},
 year = {2022}
}

@inproceedings{URGENT2024,
  title     = {{URGENT Challenge: Universality, Robustness, and Generalizability For Speech Enhancement}},
  author    = {Wangyou Zhang and Robin Scheibler and Kohei Saijo and Samuele Cornell and Chenda Li and Zhaoheng Ni and Jan Pirklbauer and Marvin Sach and Shinji Watanabe and Tim Fingscheidt and Yanmin Qian},
  year      = {2024},
  booktitle = {Interspeech 2024},
  pages     = {4868--4872},
  doi       = {10.21437/Interspeech.2024-1239},
  issn      = {2958-1796},
}

@inproceedings{URGENT2025,
  title     = {{Interspeech 2025 URGENT Speech Enhancement Challenge}},
  author    = {Kohei Saijo and Wangyou Zhang and Samuele Cornell and Robin Scheibler and Chenda Li and Zhaoheng Ni and Anurag Kumar and Marvin Sach and Yihui Fu and Wei Wang and Tim Fingscheidt and Shinji Watanabe},
  year      = {2025},
  booktitle = {{Interspeech 2025}},
  pages     = {858--862},
  doi       = {10.21437/Interspeech.2025-1363},
  issn      = {2958-1796},
}

@inproceedings{VCTK,
  title={{The Voice Bank} corpus: Design, collection and data analysis of a large regional accent speech database},
  author={Veaux, Christophe and Yamagishi, Junichi and King, Simon},
  booktitle={2013 international conference oriental COCOSDA held jointly with 2013 conference on Asian spoken language research and evaluation (O-COCOSDA/CASLRE)},
  pages={1--4},
  year={2013},
  organization={IEEE}
}

@inproceedings{LibriTTS,
  title     = {{LibriTTS}: A Corpus Derived from {LibriSpeech} for Text-to-Speech},
  author    = {Heiga Zen and Viet Dang and Rob Clark and Yu Zhang and Ron J. Weiss and Ye Jia and Zhifeng Chen and Yonghui Wu},
  year      = {2019},
  booktitle = {Interspeech 2019},
  pages     = {1526--1530},
  doi       = {10.21437/Interspeech.2019-2441},
  issn      = {2958-1796},
}

@inproceedings{WavTokenizer,
  title={{WavTokenizer}: an Efficient Acoustic Discrete Codec Tokenizer for Audio Language Modeling},
  author={Ji, Shengpeng and Jiang, Ziyue and Wang, Wen and Chen, Yifu and Fang, Minghui and Zuo, Jialong and Yang, Qian and Cheng, Xize and Wang, Zehan and Li, Ruiqi and others},
  booktitle={The Thirteenth International Conference on Learning Representations},
  year={2024}
}

@article{HiFiGAN,
  title={{HiFi-GAN}: Generative adversarial networks for efficient and high fidelity speech synthesis},
  author={Kong, Jungil and Kim, Jaehyeon and Bae, Jaekyoung},
  journal={Advances in neural information processing systems},
  volume={33},
  pages={17022--17033},
  year={2020}
}

@article{PASE,
        title={{PASE: Leveraging the Phonological Prior of WavLM for Low-Hallucination Generative Speech Enhancement}}, 
      author={Xiaobin Rong and Qinwen Hu and Mansur Yesilbursa and Kamil Wojcicki and Jing Lu},
        journal={Accepted by AAAI}, 
        year={2026}
}

@inproceedings{URGENT2025_rank1_tencent,
  title     = {{Scaling beyond Denoising: Submitted System and Findings in URGENT Challenge 2025}},
  author    = {Zhihang Sun and Andong Li and Tong Lei and Rilin Chen and Meng Yu and Chengshi Zheng and Yi Zhou and Dong Yu},
  year      = {2025},
  booktitle = {{Interspeech 2025}},
  pages     = {873--877},
  doi       = {10.21437/Interspeech.2025-795},
  issn      = {2958-1796},
}

@inproceedings{URGENT2025_rank2_nju,
  title     = {{TS-URGENet: A Three-stage Universal Robust and Generalizable Speech Enhancement Network}},
  author    = {Xiaobin Rong and Dahan Wang and Qinwen Hu and Yushi Wang and Yuxiang Hu and Jing Lu},
  year      = {2025},
  booktitle = {{Interspeech 2025}},
  pages     = {863--867},
  doi       = {10.21437/Interspeech.2025-734},
  issn      = {2958-1796},
}

@inproceedings{URGENT2025_rank3_ut,
  title     = {{FUSE: Universal Speech Enhancement using Multi‐Stage Fusion of Sparse Compression and Token Generation Models for the URGENT 2025 Challenge}},
  author    = {Nabarun Goswami and Tatsuya Harada},
  year      = {2025},
  booktitle = {{Interspeech 2025}},
  pages     = {883--887},
  doi       = {10.21437/Interspeech.2025-251},
  issn      = {2958-1796},
}

@article{URGENT2026_baseline,
  title={Less is More: Data Curation Matters in Scaling Speech Enhancement},
  author={Li, Chenda and Zhang, Wangyou and Wang, Wei and Scheibler, Robin and Saijo, Kohei and Cornell, Samuele and Fu, Yihui and Sach, Marvin and Ni, Zhaoheng and Kumar, Anurag and others},
  journal={arXiv preprint arXiv:2506.23859},
  year={2025}
}

@misc{URGENT2026,
      title={{ICASSP 2026 URGENT Speech Enhancement Challenge}}, 
      author={Chenda Li and Wei Wang and Marvin Sach and Wangyou Zhang and Kohei Saijo and Samuele Cornell and Yihui Fu and Zhaoheng Ni and Tim Fingscheidt and Shinji Watanabe and Yanmin Qian},
      year={2026},
      eprint={2601.13531},
      archivePrefix={arXiv},
      primaryClass={eess.AS},
      url={https://arxiv.org/abs/2601.13531}, 
}
%\end{small}

\end{document}